\documentclass{IEEEtran4PSCC}
\ifCLASSINFOpdf
   \usepackage[pdftex]{graphicx}
\else
   \usepackage[dvips]{graphicx}
\fi
\usepackage[cmex10]{amsmath}
\usepackage{url}
\usepackage[T1]{fontenc}
\usepackage[latin1]{inputenc}
\usepackage{eurosym}
\usepackage{caption}
\usepackage[us]{datetime} 

\usepackage[draft=False]{hyperref} 
\hypersetup{
     colorlinks=true,
     linkcolor=blue,
     filecolor=blue,
     citecolor=black,      
     urlcolor=cyan,
     }

\newcommand{\casetwostart}{\formatdate{20}{05}{2019}}
\newcommand{\casetwoend}{\formatdate{16}{06}{2019}}

\newcommand{\twelveofjune}{\formatdate{12}{06}{2019}}
\newcommand{\fiveofjune}{\formatdate{5}{06}{2019}}

\newcommand{\devices}[1]{\ensuremath{\mathcal{D}^\text{#1}}}
\newcommand{\sheddableDevices}{\ensuremath{\devices{she}}}
\newcommand{\nonflexibleDevices}{\ensuremath{\devices{nfl}}}
\newcommand{\steerableDevices}{\ensuremath{\devices{ste}}}
\newcommand{\nonsteerableDevices}{\ensuremath{\devices{nst}}}

\newcommand{\price}[1]{\ensuremath{\pi^{\text{#1}}}}
\newcommand{\curtailmentPrice}{\price{nst}}
\newcommand{\sheddingPrice}{\price{she}}
\newcommand{\steerablePrice}{\price{ste}}

\newcommand{\gridBuyPrice}{\price{i}}
\newcommand{\gridSalePrice}{\price{e}}

\newcommand{\curtail} {\ensuremath{a^\text{nst}}}
\newcommand{\steer}{\ensuremath{a^\text{ste}}}
\newcommand{\nonSteerable}{\ensuremath{P^\text{nst}}}
\newcommand{\steerable}{\ensuremath{P^\text{ste}}}

\newcommand{\shed} {\ensuremath{a^\text{she}}}
\newcommand{\nonFlexible}{\ensuremath{C^\text{nfl}}}
\newcommand{\sheddable}{\ensuremath{C^\text{she}}}

\newcommand{\BSSs}{\ensuremath{\mathcal{D}^{sto}}}
\newcommand{\SOC}{\ensuremath{s}}
\newcommand{\maxcharge}{\ensuremath{\overline{S}}}
\newcommand{\mincharge}{\ensuremath{\underline{S}}}
\newcommand{\chargerate}{\ensuremath{\overline{P}}}
\newcommand{\dischargerate}{\ensuremath{\underline{P}}}

\newcommand{\chargeEfficiency}{\ensuremath{\eta^{\text{cha}}}}
\newcommand{\dischargeEfficiency}{\ensuremath{\eta^{\text{dis}}}}
\newcommand{\initialCharge}{\ensuremath{S^\text{init}}}

\newcommand{\charge}{\ensuremath{a^\text{cha}}}
\newcommand{\discharge}{\ensuremath{a^\text{dis}}}
\newcommand{\finalCharge}{\ensuremath{S^{\text{end}}}}
\newcommand{\BSSsFee}{\ensuremath{\gamma^\text{sto}}}

\newcommand{\exportGrid}{\ensuremath{e}^\text{gri}}
\newcommand{\importGrid}{\ensuremath{i}^\text{gri}}

\newcommand{\foralltOP}{\ensuremath{\forall t' \in \mathcal{T}^m_a(t)}}
\newcommand{\foralltRTO}{\ensuremath{\forall t \in \mathcal{T}_i(t)}}

\newcommand{\forallb}{\ensuremath{\forall d \in \BSSs}}

\newcommand{\reserve}[1]{\ensuremath{r^\text{#1}}}
\newcommand{\reserveBSSInc}{\ensuremath{r^{s+}_{d,t'}}}
\newcommand{\reserveBSSDec}{\ensuremath{r^{s-}_{d,t'}}}

\newcommand{\maxExportToGrid}{\ensuremath{E}^{\text{cap}}}
\newcommand{\maxImportFromGrid}{\ensuremath{I}^{\text{cap}}}

\makeatletter
\let\old@ps@headings\ps@headings
\let\old@ps@IEEEtitlepagestyle\ps@IEEEtitlepagestyle
\def\psccfooter#1{%
    \def\ps@headings{%
        \old@ps@headings%
        \def\@oddfoot{\strut\hfill#1\hfill\strut}%
        \def\@evenfoot{\strut\hfill#1\hfill\strut}%
    }%
    \def\ps@IEEEtitlepagestyle{%
        \old@ps@IEEEtitlepagestyle%
        \def\@oddfoot{\strut\hfill#1\hfill\strut}%
        \def\@evenfoot{\strut\hfill#1\hfill\strut}%
    }%
    \ps@headings%
}
\makeatother

\psccfooter{%
        \parbox{\textwidth}{\hrulefill \\ \small{21st Power Systems Computation Conference} \hfill \begin{minipage}{0.2\textwidth}\centering \vspace*{4pt} \includegraphics[scale=0.06]{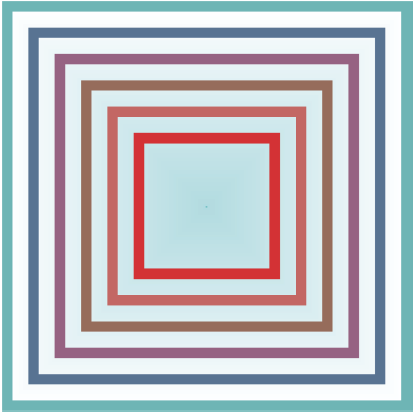}\\\small{PSCC 2020} \end{minipage} \hfill \small{Porto, Portugal --- June 29 -- July 3, 2020}}%
}

\usepackage{flushend}

\begin{document}
%
	\title{Coordination of operational planning and real-time optimization in microgrids}

	\author{
	\IEEEauthorblockN{Jonathan Dumas\IEEEauthorrefmark{1}, Selmane Dakir\IEEEauthorrefmark{1}, Cl\'ement Liu\IEEEauthorrefmark{2}, Bertrand Corn\'elusse\IEEEauthorrefmark{1}}
	\IEEEauthorblockA{\IEEEauthorrefmark{1}Department of electrical engineering and computer science\\
		ULi\`ege, Li\`ege, Belgium\\
		\{jdumas, s.dakir, bertrand.cornelusse\}@uliege.be}
	\IEEEauthorblockA{\IEEEauthorrefmark{2}clement.liu@polytechnique.edu}
}


\maketitle

	\begin{abstract}
	Hierarchical microgrid control levels range from distributed device level controllers that run at a high frequency to centralized controllers optimizing market integration that run much less frequently. Centralized controllers are often subdivided into operational planning controllers that optimize decisions over a time horizon of one or several days, and real-time optimization controllers that deal with actions in the current market period. The coordination of these levels is of paramount importance. In this paper, we propose a value function-based approach as a way to propagate information from operational planning to real-time optimization. We apply this method to an environment where operational planning, using day-ahead forecasts, optimizes at a market period resolution the decisions to minimize the total energy cost and revenues, the peak consumption and injection-related costs, and plans for reserve requirements. While real-time optimization copes with the forecast errors and yields implementable actions based on real-time measurements. The approach is compared to a rule-based controller on three use cases, and its sensitivity to forecast error is assessed.
\end{abstract}



\section{Introduction}
The hierarchical microgrid control levels divide a global microgrid control problem in time and space \cite{palizban2014microgrids}. Control levels range from distributed device level controllers that run at a high frequency to centralized controllers optimizing market integration that run much less frequently. For computation time reasons, centralized controllers are often subdivided into operational planning controllers that optimize decisions over a time horizon of one or several days but with a market period resolution, \textit{e.g.}, 15 minutes, and real-time optimization controllers that deal with actions within the current market period. The coordination of these two levels is of paramount importance to achieve the safest and most profitable operational management of microgrids.
Microgrid control and management can be achieved in several ways. Control techniques and the principles of energy-storage systems are summarized in \cite{palizban2014microgrids}. A classification of microgrid control strategies into primary, secondary, and tertiary levels is done in \cite{olivares2014trends}. The two-level approach has been intensively studied. 
A double-layer coordinated control approach, consisting of the schedule layer and the dispatch layer is adopted in \cite{jiang2013energy}. The schedule layer provides an economical operation scheme including state and power of controllable units based on the look-ahead multi-step optimization, while the dispatch layer follows the schedule layer by considering power flow and voltage limits. 
A two-stage dispatch strategy for grid-connected systems is discussed in \cite{wu2014hierarchical}, where the first stage deals with the day-ahead schedule, optimizing capital and operational cost, while the lower level handles the rescheduling of the units for few hours ahead with a time resolution of 15 min. 
A two-stage control strategy for a PV BESS-ICE (Internal Combustion Engine) microgrid is implemented in \cite{sachs2016two}. 
Discrete Dynamic Programming is used in the first layer, while the second layer problem is posed as a Boundary Value Problem. 
An approach with a high-level deterministic optimizer running at a slow timescale, 15 min, coupled to a low-level stochastic controller running at higher frequency, 1 min, is studied in \cite{cominesi2017two}.
A two-layer predictive energy management system for microgrids with hybrid energy storage systems consisting of batteries and supercapacitors is considered in \cite{ju2017two}. This approach incorporates the degradation costs of the hybrid energy storage systems.
A practical Energy Management System for isolated microgrid which considers the operational constraints of Distributed Energy Resources, active-reactive power balance, unbalanced system configuration and loading, and voltage-dependent loads is studied in \cite{solanki2018practical}. 
A two-layer mixed-integer linear programming predictive control strategy was implemented and tested in simulation and experimentally in \cite{Polimeni2019}, and \cite{moretti2019assessing} implemented a two-layer predictive management strategy for an off-grid hybrid microgrid featuring controllable and non-controllable generation units and a storage system. 

In this paper, we propose a two-layer approach with a value function to propagate information from operational planning to real-time optimization. The value function-based approach shares some similarities with the coordination scheme proposed in \cite{kumar2018stochastic}, which is based on stochastic dual dynamic programming. This paper brings new contributions:
\begin{itemize}
	\item The approach is tested by accounting for forecasting errors and high-resolution data monitored on-site corresponding to a "real-life" case.
	\item The value function approach allows to deal with indeterminacy issues. When there are several optimal solutions to the upper-level problem, this is accounted for in the lower level part, and a bias term can be added to favor one type of behavior over another, \textit{e.g.}, charge early. 
	\item This methodology is fully compatible with the energy markets as it can deal with imbalance, reserve, and dynamic selling/purchasing prices.
\end{itemize}

This paper reports results on an industrial microgrid capable of on/off grid operation. Generation and consumption forecasts are based on weather forecasts obtained with the MAR model~\cite{fettweis2017reconstructions}. 
It is organized as follows. Section \ref{sec:pb_statement} formulates the problem in an abstract manner. Section \ref{sec:proposed_method} introduces the novel two-level value function-based approach and the assumptions made. Section \ref{sec:test-description} describes the numerical tests. Section \ref{sec:numerical results} reports the results. Conclusions are drawn in Section \ref{sec:conclusion}. Section \ref{sec:notation} summarizes the notation. The methodology used for forecasting is reported as an annex.

\section{Problem statement}\label{sec:pb_statement}
A global microgrid control problem can be defined, for a given microgrid design and configuration, as operating a microgrid safely and in an economically efficient manner, by harvesting as much renewable energy as possible, operating the grid efficiently, optimizing the service to the demand side, and optimizing other side goals. We refine this definition below and start by making a few assumptions.

\subsection{Assumptions}
In this paper, the control optimizes economical criteria, which are only related to active power. All devices are supposed to be connected to the same electrical bus, which can be connected or disconnected from a public grid permanently or dynamically. 
Device-level controllers offer an interface to communicate their operating point and constraints, \textit{e.g.}, maximum charge power as a function of the current state, and implement control decisions to reach power set-points. Fast load-frequency control, islanding management, as well as reactive power control, are not in scope. The microgrid is a price taker in energy and reserve markets.

\subsection{Formulation}
Abstractly, a microgrid optimization problem can be formulated as follows
\begin{subequations}
	\begin{align}
	\min_{\mathbf{a}} \quad & \sum_{t\in \mathcal{T}_l} c(a_t, s_t, \omega_t) \\
	\text{s.t.} \ \forall t \in \mathcal{T}_l, \ &  s_{t+\Delta t} = f(a_t, s_t, \omega_t, \Delta t), \\
	& s_t \in \mathcal{S}_t .
	\end{align}
\end{subequations}
A controller has to return a set of actions $a_t = (a^m_t, a^d_t)$ at any time $t$ over the life of the microgrid ($\mathcal{T}_l$). Actions should be taken as frequently as possible to cope with the economic impact of the variability of the demand and generation sides, but not too often to let transients vanish, \textit{e.g.}, every few seconds. The time delta between action $a_t$ and the next action taken is denoted by $\Delta_t$, and is not necessarily constant.  Some of these actions are purely market-related $a^m_t$, while other actions are communicated as set-points to the devices of the microgrid $a^d_t$. The state $s_t= (s^m_t, s^d_t)$ of the microgrid at time $t$ is thus also made of two parts: $s^d_t$ represents the state of the devices, such as the state of charge of a storage system or the state of a flexible load, while $s^m_t$ gathers information related to the current market position, such as the nominated net position of the microgrid over the next market periods. The cost function $c$ gathers all the economical criteria considered. The transition function $f$ describes the physical and net position evolution of the system. 
At time instants $t \in \{\Delta_\tau, 2 \Delta_\tau, \ldots\}$, with $\Delta_\tau$ the market period, some costs are incurred based on the value of some state variables, which are then reset for the next market period. 
This problem is very difficult to solve since the evolution of the system is uncertain, actions have long-term consequences, and are both discrete and continuous. Furthermore, although functions $f$ and $c$ are assumed time-invariant, they are generally non-convex and parameterized with stochastic variables $\omega_t$. 

\section{Proposed method}\label{sec:proposed_method}
\begin{figure}[tb]
	\centering
	\includegraphics[width=0.95\linewidth]{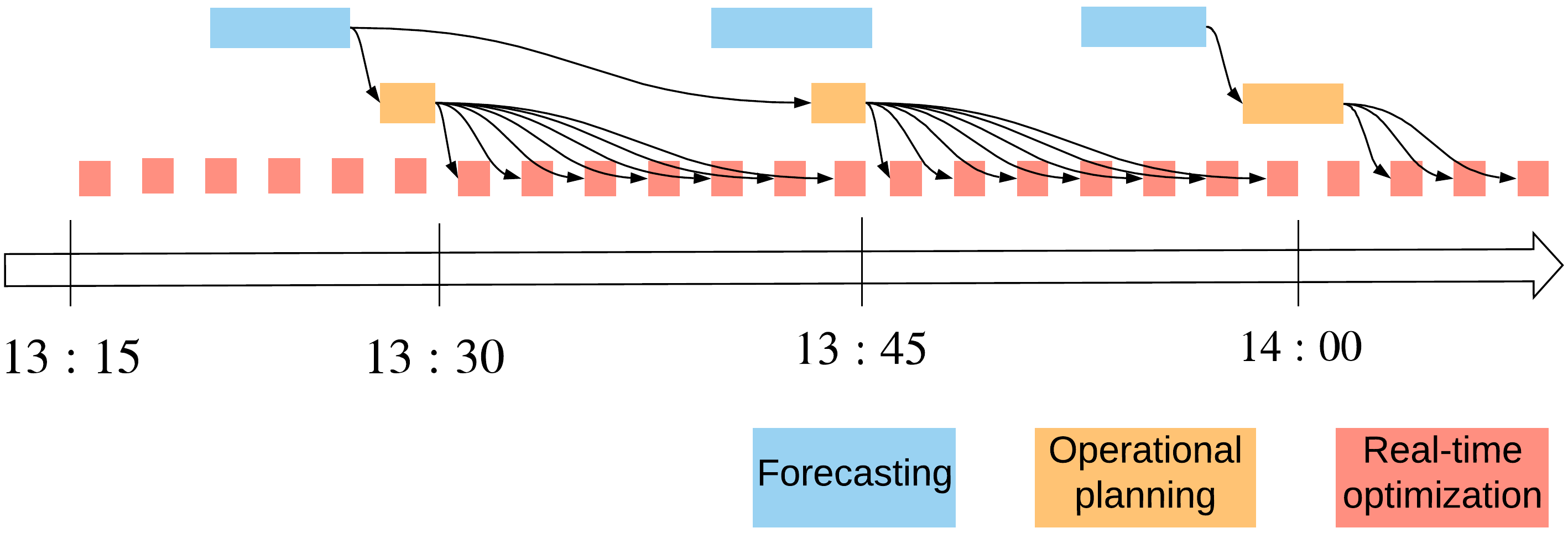}
	\caption{Hierarchical control procedure illustration.}
	\label{fig:emsschedulingpaperversion}
\end{figure}

In practice, solving the microgrid optimization problem above amounts, at every time $t$, to forecasting the stochastic variables $\boldsymbol{\omega}_{\mathcal{T}_l(t)} $, then solving the problem\footnote{Which is here expressed as a deterministic problem for simplicity, but should be treated as a stochastic problem in practice.}
\begin{subequations}
	\begin{align}
	\mathbf{a}^\star_{\mathcal{T}_l(t)}  = \arg\min \ & \sum_{t'\in \mathcal{T}_l(t)} c(a_{t'}, s_{t'}, \hat{\omega}_{t'}) \\
	\text{s.t.} \ \forall t' \in \mathcal{T}_l(t), \ & s_{t'+\Delta t'} = f(a_{t'}, s_{t'}, \hat{\omega}_{t'}, \Delta t'),\\
	& s_{t'} \in S_t' ,
	\end{align}
\end{subequations}
and applying $a^\star_t$ (potentially changing $a^{m, \star}_t$ at some specific moments only).
As forecasts are valid only for a relatively near future and optimizing over a long time horizon would anyway be incompatible with real-time operation, this problem is approximated by cropping the lookahead horizon to $\mathcal{T}_a(t) \subset \mathcal{T}_l(t)$.
However, market decisions must be refreshed much less frequently than set-points. We thus propose to further decompose the problem in an operational planning problem (OPP) for $\mathcal{T}^m_a(t)$ that computes market decisions
\begin{subequations}
	\label{eq:OPP}
	\begin{align}
	\mathbf{a}^{m, \star}_{\mathcal{T}^m_a(t)}  = \arg\min \ & \sum_{t'\in \mathcal{T}^m_a(t)} c^m(a^m_{t'}, s_{t'}, \hat{\omega}_{t'}) \\
	\text{s.t.} \ \forall t' \in \mathcal{T}^m_a(t),  \ & s_{t'+\Delta \tau} = f^m(a^m_{t'}, s_{t'}, \hat{\omega}_{t'}, \Delta \tau) \\
	& s_{t'} \in S_{t'} ,
	\end{align}
\end{subequations}
and a real-time problem (RTP) that computes set-points for time $t$
\begin{subequations}
	\label{eq:RTP}
	\begin{align}
	a^{d, \star}_{t}  = \arg\min \ & c^d(a^d_{t}, s_{t}, \hat{\omega}_{t}) + v_{\tau(t)}(s_{\tau(t)}) \\
	\text{s.t.} \ & s_{\tau(t)} = f^d(a^d_{t}, s_{t}, \hat{\omega_{t}}, \tau(t) - t)\\
	& s_{\tau(t)} \in S_{\tau(t)} ,
	\end{align}
\end{subequations}
with $c(a_{t}, s_{t}, w_{t}) = c^m(a^m_{t}, s_{t}, w_{t}) + c^d(a^d_{t}, s_{t}, \omega_{t})$. 
The function $v_t$ is the cost-to-go as a function of the state of the system at the end of the ongoing market period, it regularizes decisions of RTP to account for the longer-term evolution of the system.
We detail hereunder how we obtain $v_t$.  An overview of the approach is depicted in Figure~\ref{fig:emsschedulingpaperversion}.

\subsection{Computing the cost-to-go function $v_{\tau(t)}$}
The function $v_t$ represents the optimal value of (\ref{eq:OPP}) as a function of the initial state $s_{\tau(t)}$ of this problem. If we make the assumption that (\ref{eq:OPP}) is modeled as a linear program, the function $v_{\tau(t)}$ is thus convex and piecewise linear. Every evaluation of (\ref{eq:OPP}) with the additional constraint\footnote{The $\perp \mu$ notation means that $\mu$ is the dual variable of the constraint.}
\begin{align}
s_{\tau(t)} = s' \perp \mu
\end{align}
yields the value $v_{\tau(t)}(s')$ and a supporting inequality (a cut)
\begin{align}
v_{\tau(t)}(s) \geq v_{\tau(t)}(s') + \mu^T  s.
\end{align}
The algorithm to approximate $v_{\tau(t)}(s')$ works as follows:
\begin{enumerate}
	\item estimate the domain of $v_{\tau(t)}$, \textit{i.e.}, the range of states reachable at time $\tau(t)$ and the most probable state that will be reached $s^\star_{\tau(t)}$;
	\item evaluate $v_{\tau(t)}(s^\star_{\tau(t)})$ and the associated $\mu^\star$;
	\item repeat step 2 for other state values until all regions of $v_{\tau(t)}$ are explored. 
\end{enumerate}
Note that if the state is of dimension one and  (\ref{eq:OPP}) is a linear program, simplex basis validity information can be used to determine for which part of the domain of $v_{\tau(t)}$ the current cut is tight, else a methodology such as proposed in \cite{bemporad2003greedy} can be used.


\subsection{OPP formulation}

The OPP objective function implemented for the case study is
\begin{align}	
	J_{\mathcal{T}^m_a(t)}^{OP} = \ & \sum_{t'\in \mathcal{T}^m_a(t)} \bigg ( C_{t'}^{OP} + D_{t'}^{OP} \bigg )
\end{align}
with Operational Planner (OP) the name of this planer. $\mathcal{T}^m_a(t)$ is composed of 96 values with $\Delta_\tau = 15 $ minutes and $T_a = 24$ hours. $C_{t'}^{OP}$ models the immediate costs and $D_{t'}^{OP}$ the delayed costs at  $t'$.
$C_{t'}^{OP}$ takes into account different revenues and costs related to energy flows: the costs of shed demand, steered and non steered generation, the revenues from selling energy to the grid, the costs of purchasing energy from the grid and the costs for using storage
\begin{align}\label{eq:OP_immediate_cost}
	C_{t'}^{OP} = \ & \sum_{t'\in \mathcal{T}^m_a(t)} \bigg (  \sum_{d\in \sheddableDevices} \Delta_\tau  \sheddingPrice_{d,t'}  \sheddable_{d,t'}  \shed_{d,t'}  \notag\\
	&+ \sum_{d\in \steerableDevices} \Delta_\tau  \steerablePrice_{d,t'}   \steerable_{d,t'}  \steer_{d,t'} \notag\\
	&+ \sum_{d\in \nonsteerableDevices} \Delta_\tau  \curtailmentPrice_{d,t'}  \nonSteerable_{d,t'}  \curtail_{d,t'} \notag \\ 
	&+ \sum_{d\in \BSSs} \Delta_\tau  \BSSsFee_{d}  \left(\chargerate_{d}  \chargeEfficiency_{d}  \charge_{d,t'} + \frac{\dischargerate_d}{\dischargeEfficiency_d}  \discharge_{d,t'} \right)   \notag  \\
	& - \gridSalePrice_{t'}  \exportGrid_{t'} + \gridBuyPrice_{t'}  \importGrid_{t'} \bigg ) .
	\end{align}
$D_{t'}^{OP}$  is composed of the peak cost and symmetric reserve revenue
\begin{align}	\label{eq:OP_delayed_cost}
	D_{t'}^{OP} = \ & \price{p}  \delta p_{t'} - \price{s}_{OP}  \reserve{sym}_{t'} , 
\end{align}
$\delta p_{t'}$ is the peak difference between the previous maximum historic peak $p_h$ and the current peak within the market period $t'$. $\reserve{sym}_{t'}$ is the symmetric reserve provided to the grid within the current market period $t'$.

\subsection{OP constraints}
The first set of constraints defines bounds on state and action variables, $\foralltOP$
\begin{subequations}
	\label{eq:OP_action_constraint_set}	
	\begin{align}
	&a^k_{d,t'} \leq 1     &&\forall d \in \mathcal{D}^k, \forall k \in \{\text{ste}, \text{she}, \text{nst}\}   \\
	&\charge_{d,t'} \leq 1    &&\forallb       \\
	&\discharge_{d,t'} \leq 1 &&\forallb    \\
	&\mincharge_d \leq \SOC_{d,t'} \leq \maxcharge_d &&\forallb .
	\end{align}
\end{subequations}
The energy flows are constrained, $\foralltOP$, by
\begin{subequations}
	\label{eq:OP_energy_flows_constraint_set}	
	\begin{align}
	&(\exportGrid_{t'} - \importGrid_{t'})/\Delta_\tau
	-  \sum_{d\in \nonsteerableDevices} (1-\curtail)  \nonSteerable_{d,t'} + \sum_{d\in \steerableDevices} \steer_{d,t'}  \steerable_{d,t'} \notag\\
	& + \sum_{d\in\nonflexibleDevices} \nonFlexible_{d,t'}
	+ \sum_{d\in \sheddableDevices}(1 - \shed_{d,t'})  \sheddable_{d,t'}  \notag\\
	&+  \sum_{d\in\BSSs} \left(\chargerate_d  \charge_{d,t'} - \dischargerate_d  \discharge_{d,t'} \right)= 0 \\
	&(\exportGrid_{t'} - \importGrid_{t'})/ \Delta_\tau  \leq \maxExportToGrid_{t'} \\
	&(\importGrid_{t'} - \exportGrid_{t'})/ \Delta_\tau \leq \maxImportFromGrid_{t'} .
	\end{align}
\end{subequations}
The dynamics of the state of charge are, $\forallb$
\begin{subequations}
	\label{eq:OP_soc_dynamic}
	\begin{align}
	&s_{d,1} - \Delta_\tau \left(\chargerate_{d}  \chargeEfficiency_{d}  \charge_{d,1} - \frac{\dischargerate_{d}}{\dischargeEfficiency_d}  \discharge_{d,1} \right) = \initialCharge_{d}\\
	&s_{d,t'} - s_{d, t'-\Delta_\tau} - \Delta_\tau \left(\chargerate_{d}  \chargeEfficiency_{d}  \charge_{d,t'} - \frac{\dischargerate_{d}}{\dischargeEfficiency_d}  \discharge_{d,t'} \right)=0 \notag&&&\\
	&\hspace{4em}, \foralltOP  \\
	&s_{d,\tau(t+T_a)} = \finalCharge_{d} .
	\end{align}
\end{subequations}
The set of constraints related to the peak power $\foralltOP$
\begin{subequations}
	\label{eq:OP_peak_constraints}
	\begin{align}
	(\importGrid_{t'} - \exportGrid_{t'})/ \Delta_\tau &\leq p_{t'}  \\
	- \delta p_{t'} &\leq 0 \\
	- \delta p_{t'} &\leq - (p_{t'} - p_h) .
	\end{align}
\end{subequations}
The last constraints define symmetric reserve $\foralltOP$
\begin{subequations}
	\label{eq:OP_reserve_constraints}
	\begin{align}
	&\reserveBSSInc \leq \dfrac{\left( \SOC_{d,t'} - \mincharge_{d}\right)\dischargeEfficiency_d}{\Delta_\tau} \quad  \forallb  \\
	&\reserveBSSInc \leq \dischargerate_{d} (1-\discharge_{d,t'}) \quad  \forallb \quad   \\
	&\reserveBSSDec \leq \dfrac{\left( \maxcharge_{d} - \SOC_{d,t'} \right)}{\chargeEfficiency_d \Delta_\tau} \quad \forallb  \\
	&\reserveBSSDec \leq \chargerate_{d}(1-\charge_{d,t'}) \quad \forallb  \\
	&\reserve{sym, OP} \leq  \sum_{d\in\BSSs} \reserveBSSInc + \sum_{d\in \steerableDevices}\steerable_{d,t'} (1 - \steer_{d,t'})\notag\\
	&\hspace{4em} + \sum_{d\in \nonsteerableDevices} \nonSteerable_{d,t'} (1-\curtail)  \\
	&\hspace{4em} +\sum_{d\in \sheddableDevices} \sheddable_{d,t'} (1-\shed_{d,t'})   \notag\\
	&\reserve{sym, OP} \leq \sum_{d\in\BSSs} \reserveBSSDec + \sum_{d\in \steerableDevices} \steerable_{d,t'} \steer_{d,t'} \notag\\
	&\hspace{4em} +\sum_{d\in \nonsteerableDevices} \nonSteerable_{d,t'} \curtail +\sum_{d\in \sheddableDevices} \sheddable_{d,t'} \shed_{d,t'} .
	\end{align}
\end{subequations}

\subsection{RTP formulation}

The RTP objective function implemented for the case study is
\begin{align}\label{eq:RTO_objective_function}
	J_{t}^{RTO} = \ & C_{t}^{RTO} + D_{t}^{RTO} + v_{\tau(t)}(s_{\tau(t)})
\end{align}
with Real-Time Optimizer (RTO) the name of this controller. $C_{t}^{RTO}$ models the immediate costs, $D_{t}^{RTO}$ the delayed costs and $v_{\tau(t)}(s_{\tau(t)})$ the cost-to-go function of the state of the system at time $t$ within a current market period. 
$C_{t}^{RTO}$ is the same as $C_{t'}^{OP}$ by replacing $t'$ by $t$, $\Delta_\tau$ by $\Delta_t$ and considering only one period of time $t$.
$D_{t}^{RTO}$ is composed of the peak cost and symmetric reserve penalty costs
\begin{align}\label{eq:RTO_delayed_cost}
	D_{t}^{RTO} = \ & \price{p}  \delta p_{\tau(t-\Delta_\tau),\tau(t)} + s^{TSO}_{t} \price{s}_{RTO}  \Delta \reserve{sym} ,
	\end{align}
$\delta p_{\tau(t-\Delta_\tau),\tau(t)}$ is the peak difference between the previous maximum historic peak $p_h$ and the current peak within the market period computed by RTO. The difference with OP relies on its computation as at $t$ the market period is not finished. Thus the peak within this market period is computed by adding the peak from the beginning of the market period to $t$ and the one resulting from the actions taken from $t$ to the end of the market period. $\Delta \reserve{sym}$ is the difference between the symmetric reserve computed by OP and the current reserve within the market period computed by RTO. $s^{TSO}_{t}$ is the reserve activation signal to activate the tertiary symmetric reserve. It is set by the TSO, 0 if activated, else 1. The activation occurs at the beginning of the next market period.

\subsection{RTO constraints}
The set of constraints that defines the bounds on state and action variables and the energy flows are the same as the OP (\ref{eq:OP_action_constraint_set}) and (\ref{eq:OP_energy_flows_constraint_set}) by replacing $t'$ by $t$, $\Delta_\tau$ by $\Delta_t$ and considering only one period of time $t$.
The next constraint describes the dynamics of the state of charge $ \forallb$ and $ \foralltRTO$
\begin{align}
&s_{d,\tau(t)} - \Delta_t \left(\chargerate_{d}  \chargeEfficiency_{d}  \charge_{d,t} - \frac{\dischargerate_{d}}{\dischargeEfficiency_d}  \discharge_{d,t} \right) = \initialCharge_{d, t} .
\end{align}
The set of constraints related to the peak power $\foralltRTO$
\begin{subequations}
	\label{eq:RTO_peak_constraints}
	\begin{align}
	(\importGrid_t - \exportGrid_t)/ \Delta_t &\leq p_{t,\tau(t)}  \\
	- \delta p_{\tau(t)} &\leq 0   \\
	- \delta p_{\tau(t)} &\leq - (p_{\tau(t-\Delta_\tau),\tau(t)} - p_h)   \\
	p_{\tau(t-\Delta_\tau),\tau(t)} &= \beta  p_{\tau(t-\Delta_\tau),t} + (1-\beta )  p_{t,\tau(t)}  
	\end{align}
\end{subequations}
with $\beta = 1 - \Delta_t / \Delta_\tau$.
The last set of constraints defining the symmetric reserve are the same as the OP (\ref{eq:OP_reserve_constraints}) by replacing $t'$ by $t$, $\reserve{sym, OP}$ by $\reserve{sym, RTO}$ and adding $\foralltRTO$
\begin{subequations}
	\label{eq:RTO_reserve_constraints}
	\begin{align}
	& - \Delta \reserve{sym} \leq 0  \\
	& - \Delta \reserve{sym} \leq - (\reserve{sym, OP} - \reserve{sym, RTO}) .
	\end{align} 
\end{subequations}

\section{Test description}\label{sec:test-description}

Our case study is based on the MiRIS microgrid located at the John Cockerill Group's international headquarters in Seraing, Belgium\footnote{\url{https://johncockerill.com/fr/energy/stockage-denergie/}}. It is composed of PV, several energy storage devices, and a non-sheddable load. The load and PV data we use come from on-site monitoring. All data, including the weather forecasts, are available on the Kaggle platform\footnote{\url{https://www.kaggle.com/jonathandumas/liege-microgrid-open-data}}. 
The case study consists of comparing $\mbox{RTO-OP}$ to a Rule-Based Controller (RBC) for three configurations of the installed PV capacity, cf. Table~\ref{tab:simulation_parameters}. The RBC prioritizes the use of PV production for the supply of the electrical demand. If the microgrid is facing a long position, it charges the battery. And if this one is fully charged it exports to the main grid. If the microgrid is facing a short position it prioritizes the use of the battery to supply the demand. And if this one is fully discharged it imports from the main grid. This controller does not take into account any future information, \textit{e.g.}, PV, consumption forecasts, energy prices, or market information such as the peak of the symmetric reserve.
Case 3 is the result of a sizing study that defined the optimal device sizes given the PV and consumption data. The sizing methodology used is described in \cite{dakir2019sizing}.

\begin{figure}[tb]
	\centering
	\includegraphics[width=1\linewidth]{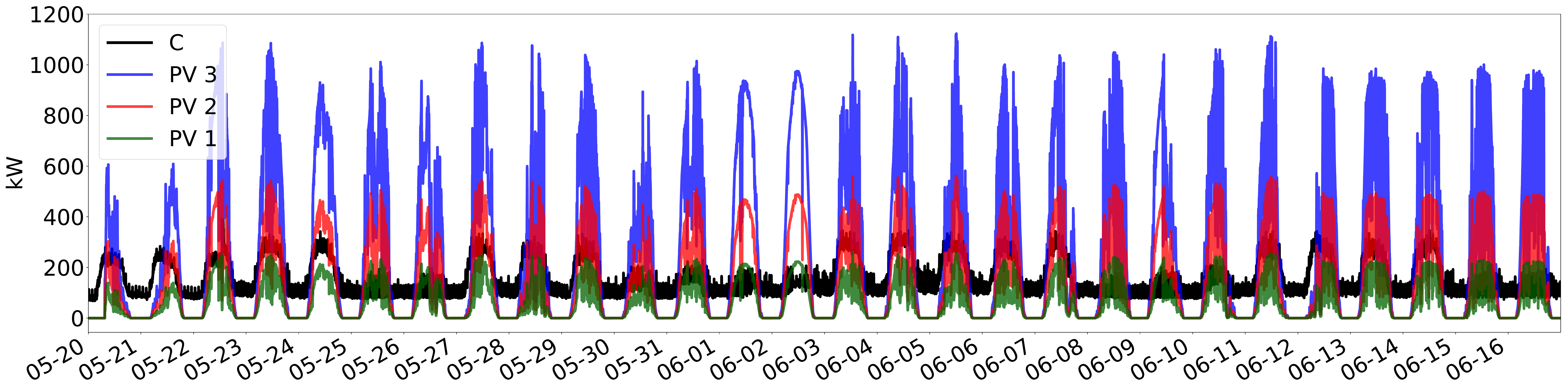} \\
	\includegraphics[width=1\linewidth]{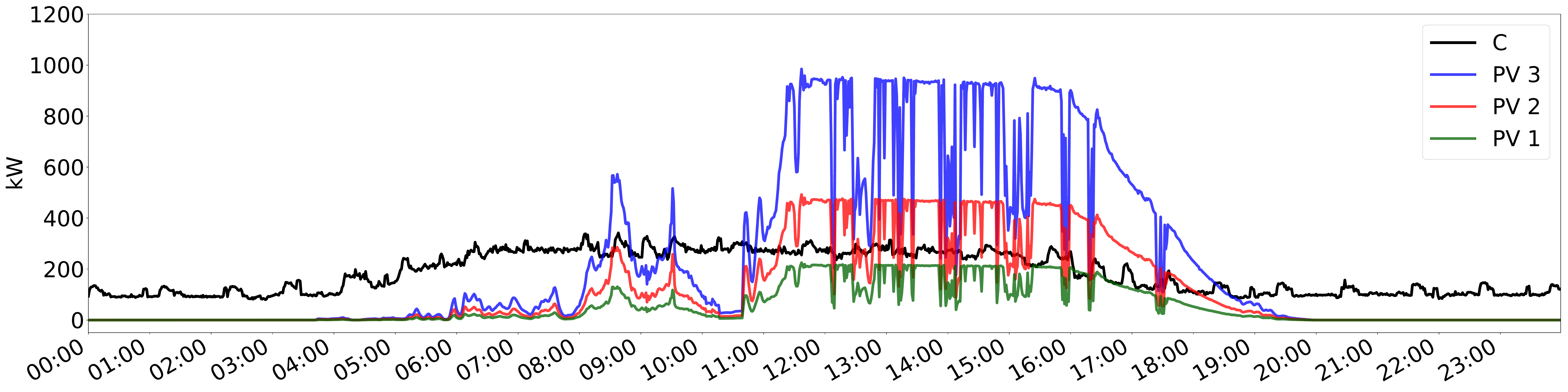}
	\captionsetup{justification=centering}
	\caption{Top: PV \& consumption simulation data. Bottom: zoom on $\twelveofjune$.}
	\label{fig:energy_data}
\end{figure}

Figure~\ref{fig:energy_data} shows the PV \& consumption data over the simulation period: from $\casetwostart$ to $\casetwoend$. The selling price $\gridSalePrice$ is constant, the purchasing price is composed of a day $\gridBuyPrice_d$ and night prices $\gridBuyPrice_n$. Day prices apply from 5 a.m. to 8 p.m. (UTC) during the weekdays and night prices apply from 8 p.m. to 5 a.m. during weekdays and the entire weekend. The peak mechanism is taken into account with a constant peak price $\price{p}$ and an initial maximum historic peak $p_h$. Storage systems are initially fully charged.
The PV and consumption data have a 1-second resolution, meaning the RTO could compute its optimization problem each five to ten seconds in operational mode. CPLEX 12.9 is used to solve all the optimization problems, on an Intel Core i7-8700 3.20 GHz based computer with 12 threads and 32 GB of RAM. The average computation time per optimization problem composed of the OP and RTO is a few seconds. However, to maintain a reasonable simulation time RTO is called every minute. The dataset is composed of 28 days with an average computation time of two hours to solve 1440 optimization problems per day, with one-minute resolution, leading to a total of two days for the entire dataset. The OP computes a planning quarterly corresponding to the Belgian market period. The computation time of the RTO on a regular computer is around a few seconds and the OP around twenty seconds. In total, the simulation computation time is up to a few hours. The OP computes quarterly planning based on PV and consumption twenty-four ahead forecasts. The weather-based forecast methodology is described in detail in Annex~\ref{sec:forecasting}. Two "classic" deterministic techniques are implemented, a Recurrent Neural Network (RNN) with the Keras Python library \cite{chollet2015keras} and a Gradient Boosting Regression (GBR) with the Scikit-learn Python library \cite{scikit-learn}. These models use as input the weather forecasts provided by the Laboratory of Climatology of the Li\`ege University, based on the MAR regional climate model \cite{fettweis2017reconstructions}. It is an atmosphere model designed for meteorological and climatic research, used for a wide range of applications, from km-scale process studies to continental-scale multi-decade simulations.
To estimate the impact of the PV and consumption forecast errors on the controllers, the simulation is performed with the OP having access to the PV and consumption future values ($\mbox{RTO-OP}^{\star}$).
Then, the simulation is performed with the symmetric reserve mechanisms to cope with the forecast errors. A constant symmetric reserve price $\price{s}_{OP}$ for the OP and a penalty reserve $\price{s}_{RTO}$ for the RTO are set to 20 (\euro / kW).

\begin{table}[tb]
	\begin{center}
		\scriptsize
		\renewcommand\arraystretch{1.5}
		\caption{Case studies parameters and data statistics.}
		\begin{tabular}{lrrrrr}
			\hline  \hline
			Case & $\mbox{PV}_p$ &  $\overline{\mbox{PV}}$ & $\mbox{PV}_{max}$ & $\mbox{PV}_{min}$ & $\mbox{PV}_{std}$ \\ \hline 
			1	& 400  &  61 & 256  & 0 & 72\\ \hline 
			2	& 875  &  133 & 561 & 0 & 157 \\ \hline 
			3	& 1750 & 267 & 1122 & 0 & 314\\ \hline  \hline
			Case & $\mbox{C}_p$ &  $\overline{\mbox{C}}$ & $\mbox{C}_{max}$ & $\mbox{C}_{min}$ & $\mbox{C}_{std}$ \\ \hline 
			1 - 3	& 1000 & 153 & 390 & 68 & 72 \\ \hline  \hline
			Case & $\mbox{S}_p$ &  $\maxcharge$, $\mincharge$ & $\dischargerate$, $\chargerate$ & $\chargeEfficiency$, $\dischargeEfficiency$ & $\initialCharge$ \\ \hline 
			1 - 3 & 1350 & 1350, 0 & 1350, 1350 & 0.95, 0.95 & 100 \\ \hline  \hline
			Case & $p_h$, $\price{p}$ &  $\maxImportFromGrid$ & $\maxExportToGrid$ & $\gridBuyPrice_d$, $\gridBuyPrice_n$ & $\gridSalePrice$ \\ \hline 
			1 - 3	& 150, 40 & 1500 & 1500 & 0.2, 0.12 & 0.035 \\ \hline \hline 
		\end{tabular}
		\label{tab:simulation_parameters}
	\end{center}
\end{table}

\section{Numerical results}\label{sec:numerical results}

\subsection{No symmetric reserve}
Table \ref{tab:results_wht_reserve} provides the simulation results without taking into account the symmetric reserve. The smaller the PV installed capacity the higher the peak and energy costs. The $\mbox{RTO-OP}^{\star}$ provides the minimal peak cost whereas the RBC provides the minimal energy cost on all cases. However, $\mbox{RTO-OP}^{\star}$ achieves the minimal total cost, composed of the energy and peak costs.
\begin{table}[tb]
	\begin{center}
		\scriptsize
		\renewcommand\arraystretch{1.5}
		\caption{Results without symmetric reserve.}
		\begin{tabular}{lrrrrrr}
			\hline  \hline 
			Case 1 & $c_E$ & $c_p$ & $c_t$ & $\Delta_p$ & $\mbox{I}_{tot}$ & $\mbox{E}_{tot}$ \\ \hline  
			RBC	                     & 10.13 & 6.68  & 16.81 & 167 & 61 &  0    \\ \hline 
			$\mbox{RTO-OP}^{RNN}$	 & 10.37 & 3.62  & 13.99 & 91  & 64 & 1    \\ \hline 
			$\mbox{RTO-OP}^{GBR}$	 & 10.25 & 5.27  & 15.53 & 132 & 63 & 1    \\ \hline 
			$\mbox{RTO-OP}^{\star}$	 & 10.24 & 0.99  & 11.23 & 25  & 64 & 1    \\ \hline  \hline 
			Case 2 & $c_E$ & $c_p$ & $c_t$ & $\Delta_p$ & $\mbox{I}_{tot}$ & $\mbox{E}_{tot}$ \\ \hline  
			RBC	                     & 3.19  & 4.85  & 8.04 & 121 & 22 &  7  \\ \hline 
			$\mbox{RTO-OP}^{RNN}$	 & 4.78  & 2.87  & 7.65 & 72  & 31 & 15   \\ \hline 
			$\mbox{RTO-OP}^{GBR}$	 & 4.30  & 4.90  & 9.2  & 123 & 28 & 13 \\ \hline 
			$\mbox{RTO-OP}^{\star}$	 & 4.06  & 0     & 4.06 & 0   & 26 & 10\\ \hline \hline 
			Case 3 & $c_E$ & $c_p$ & $c_t$ & $\Delta_p$ & $\mbox{I}_{tot}$ & $\mbox{E}_{tot}$ \\ \hline  
			RBC	                     & -2.13 & 4.12  & 1.99 & 105   & 3 & 77  \\ \hline 
			$\mbox{RTO-OP}^{RNN}$	 & -1.66 & 4.12  & 2.46 & 105   & 7 & 80   \\ \hline 
			$\mbox{RTO-OP}^{GBR}$	 & -1.67 & 4.23  & 2.56 & 106   & 7 & 81 \\ \hline 
			$\mbox{RTO-OP}^{\star}$	 & -1.90 & 0     & 0 & 0     & 5 & 79 \\ \hline  \hline 
		\end{tabular}
		\label{tab:results_wht_reserve}
	\end{center}
\end{table}
This simulation illustrates the impact of the forecasts on the $\mbox{RTO-OP}$ behavior. The RNN forecaster provides the best results but the $\mbox{RTO-OP}^{RNN}$ is still a long way to manage the peak as $\mbox{RTO-OP}^{\star}$ due to the forecasting errors. The peak cost strongly penalizes the benefits as it applies on all the year ahead once it has been reached. 

\begin{figure}[tb]
	\centering
	\includegraphics[width=1\linewidth]{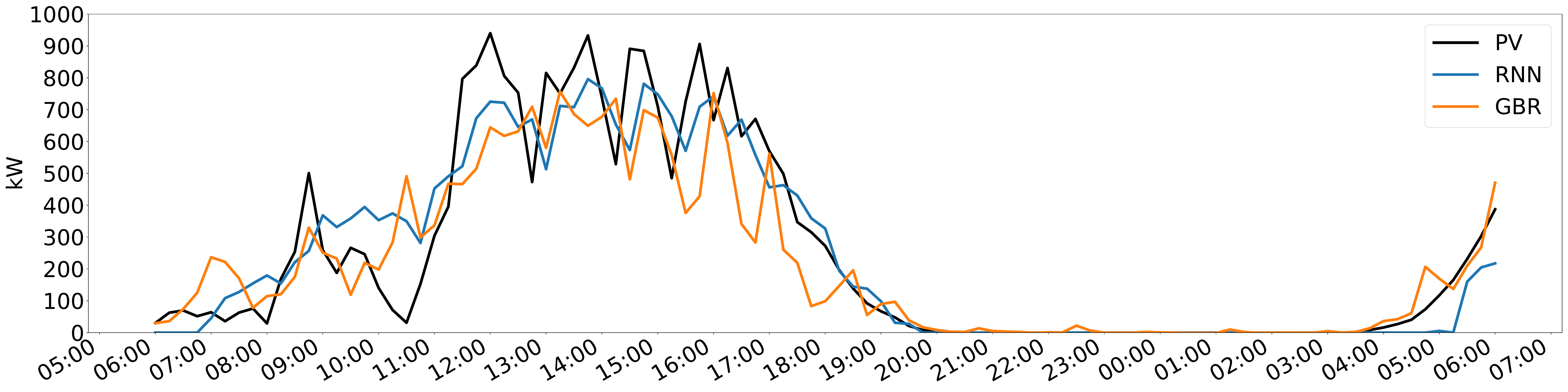}  %
	\captionsetup{justification=centering}
	\caption{Case 3 PV forecast on $\twelveofjune$, 06h00 UTC.}
	\label{fig:case3_pv_forecast_12062019}
\end{figure}
\begin{figure}[tb]
	\centering
	\includegraphics[width=1\linewidth]{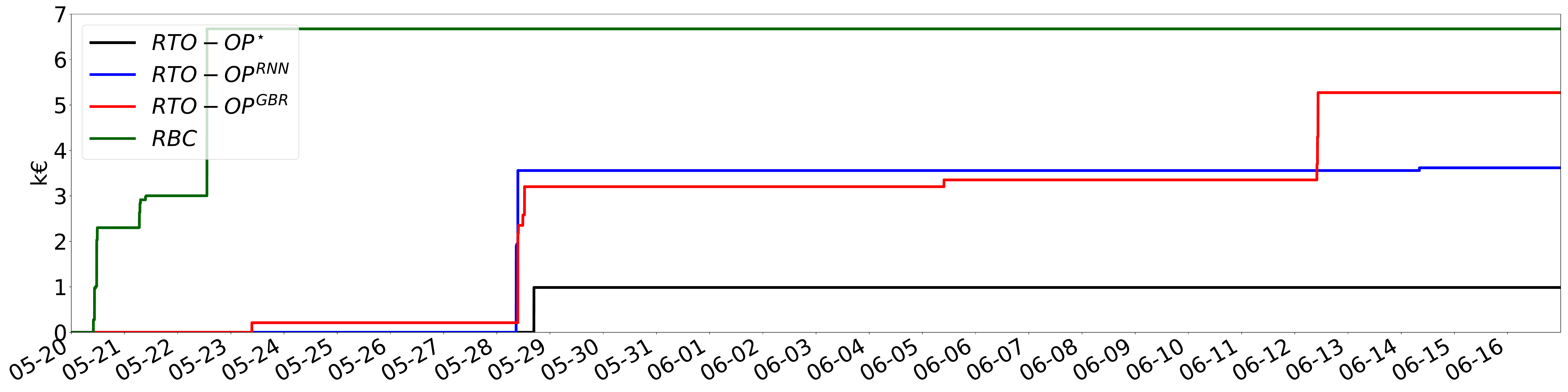} \\[1mm] 
	\includegraphics[width=1\linewidth]{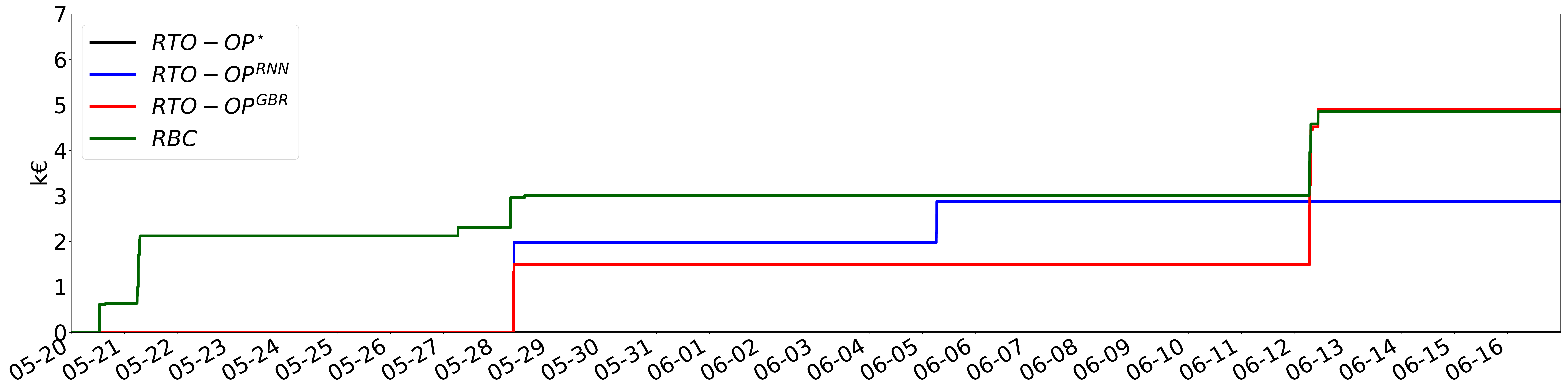} \\[1mm] 
	\includegraphics[width=1\linewidth]{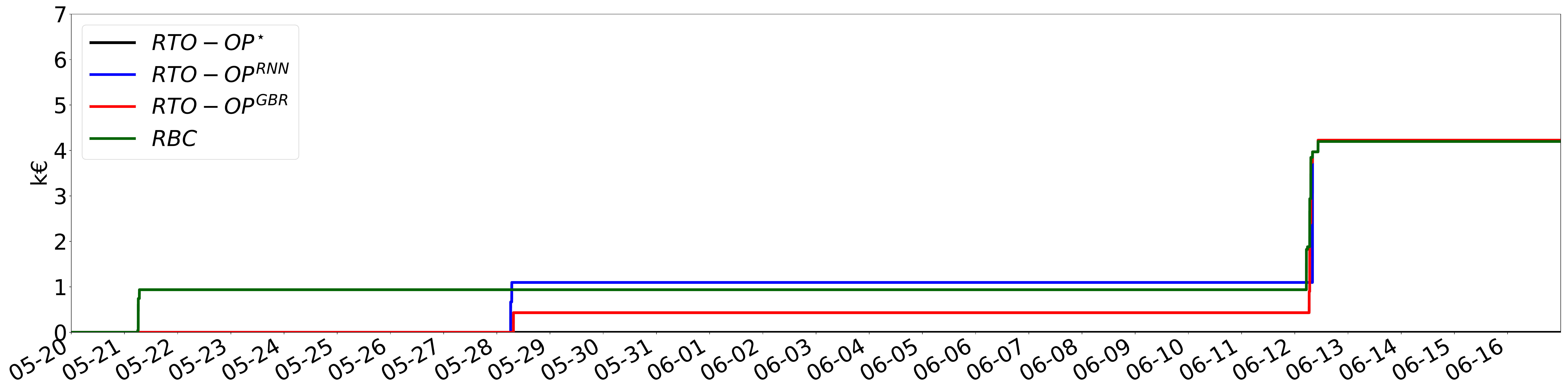} 
	\captionsetup{justification=centering}
	\caption{Case 1 (top), 2 (middle), 3 (bottom) cumulative peak costs.}
	\label{fig:case321_peak_cost}
\end{figure}
\begin{figure}[tb]
	\centering
	\includegraphics[width=0.95\linewidth]{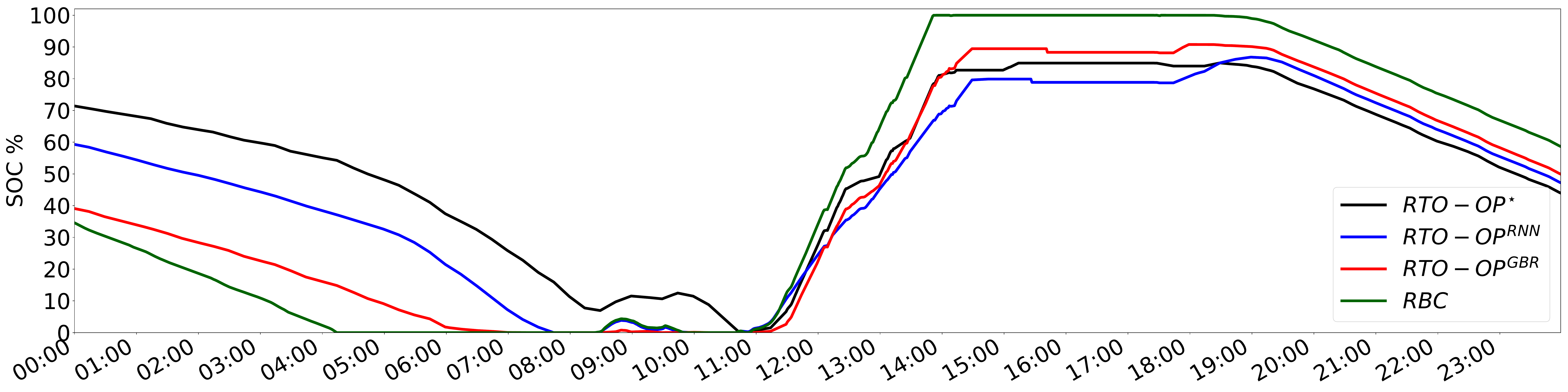}  \\[1mm] 
	\includegraphics[width=1\linewidth]{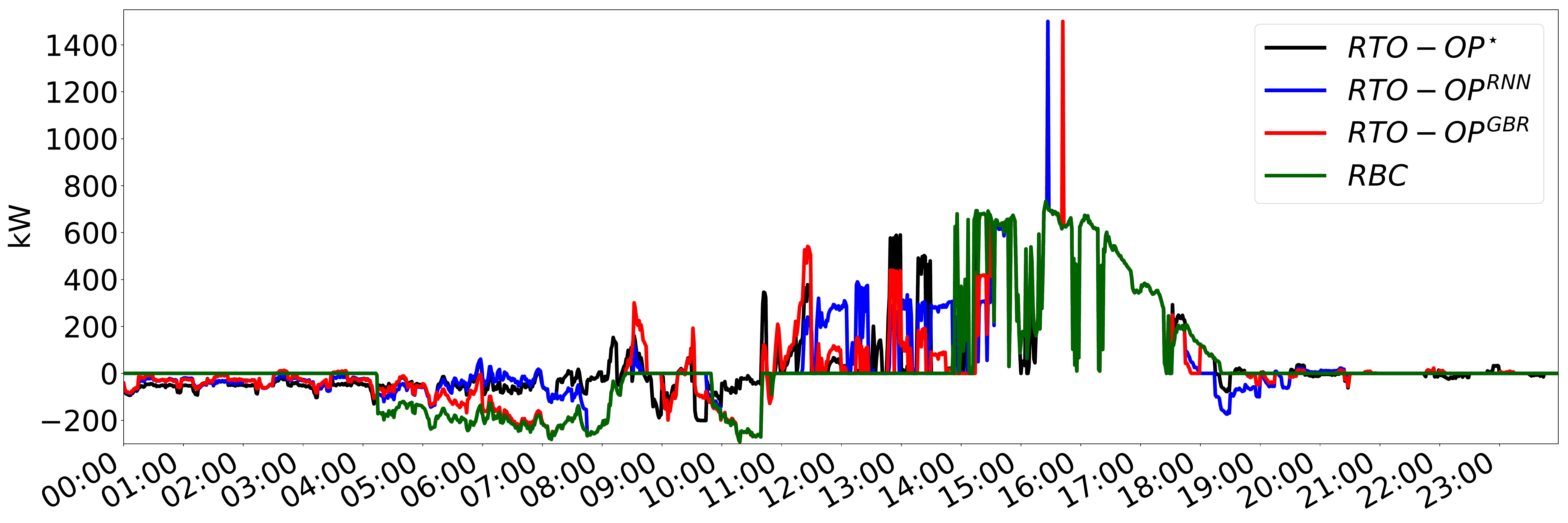} 
	\captionsetup{justification=centering}
	\caption{Case 3 SOC (top) and net export power (bottom) on $\twelveofjune$.}
	\label{fig:case3_soc_net_1206}
\end{figure}

In case 3, all the controllers except $\mbox{RTO-OP}^{\star}$ reached the maximum peak on $\twelveofjune$ around 10:30 a.m. as shown on Figure~\ref{fig:case321_peak_cost}. Figure~\ref{fig:energy_data} shows a sudden drop in the PV production around 10 a.m. that is not accurately forecasted by the RNN and GBR forecasters as shown in Figure~\ref{fig:case3_pv_forecast_12062019}. This prediction leads to a non accurate planning of OP. Thus, the RTO cannot anticipate this drop and has to import at the last minute energy to balance the microgrid. Figure~\ref{fig:case3_soc_net_1206} shows the controllers behavior on $\twelveofjune$ where the peak is reached.
In case 2 all controllers reached the same peak as in case 3 except $\mbox{RTO-OP}^{RNN}$ that reached a smaller one on $\fiveofjune$. The forecasts accuracy explains this behavior as in case 3.
Finally in case 1, each controller reached a different peak. The smallest one is achieved by the $\mbox{RTO-OP}^{\star}$, followed by the $\mbox{RTO-OP}^{RNN}$.
These cases show that the $\mbox{RTO-OP}$ controller optimizes PV-storage usage, and thus requires less installed PV capacity for a given demand level. This result was expected as the peak management is not achieved by the RBC and becomes critical when the PV production is smaller than the consumption. This simulation also demonstrates the forecast accuracy impact on the $\mbox{RTO-OP}$ behavior. 

\subsection{Results with symmetric reserve}

\begin{figure}[tb]
	\centering
	\includegraphics[width=0.9\linewidth]{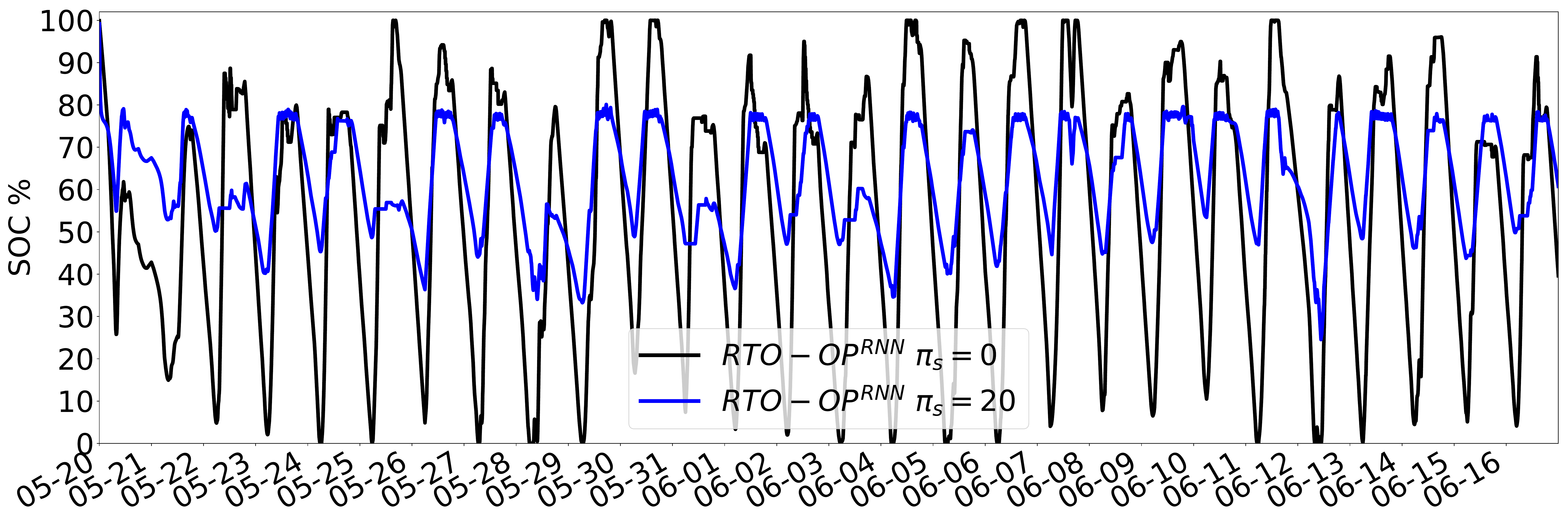}  
	\caption{Case 3 SOC comparison for $\mbox{RTO-OP}^{RNN}$ with and without symmetric reserve.}
	\label{fig:case3_soc_r0_vs_20}
\end{figure}

\begin{figure}[tb]
	\centering
	\includegraphics[width=0.9\linewidth]{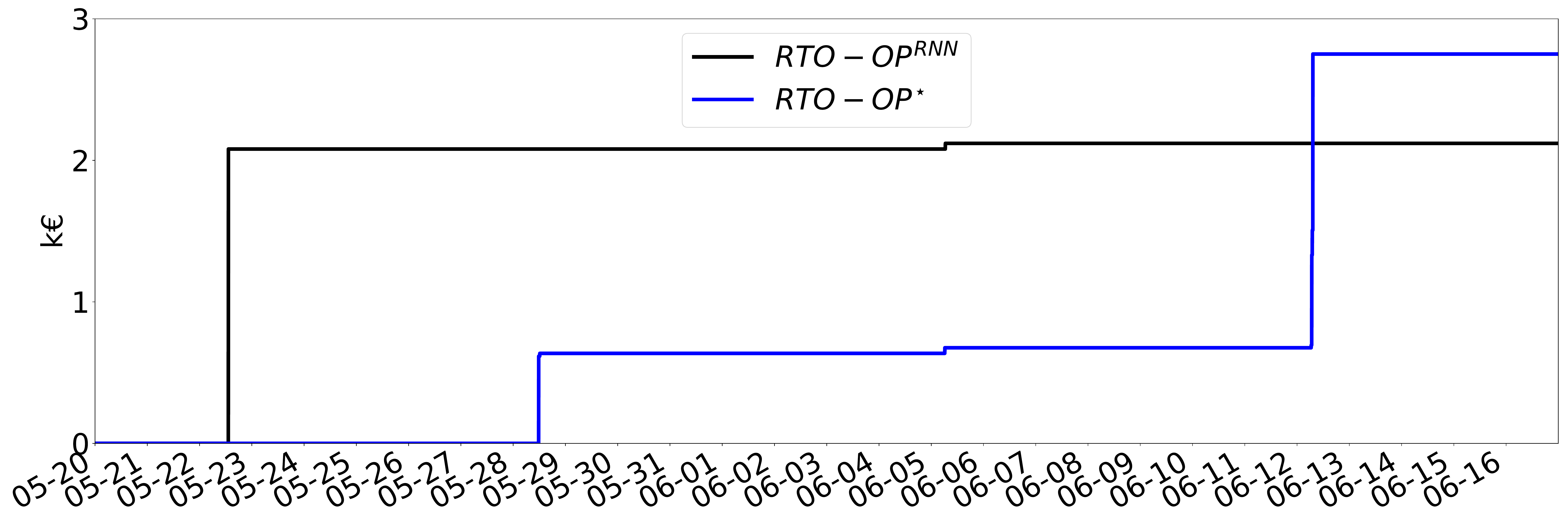}  \\[1mm] 
	\includegraphics[width=0.9\linewidth]{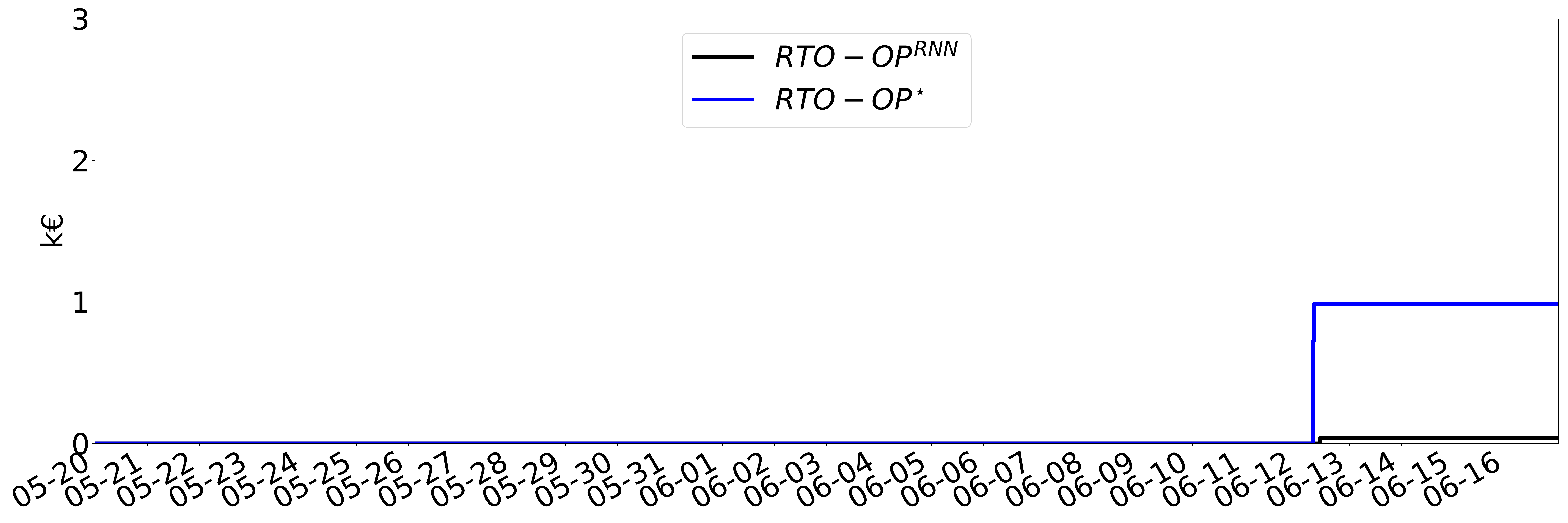} 
	\captionsetup{justification=centering}
	\caption{Case 1 (top) and 2 (bottom) cumulative peak costs.}
	\label{fig:case21_with_reserve_peak_costs}
\end{figure}

\begin{figure}[tb]
	\centering
	\includegraphics[width=0.9\linewidth]{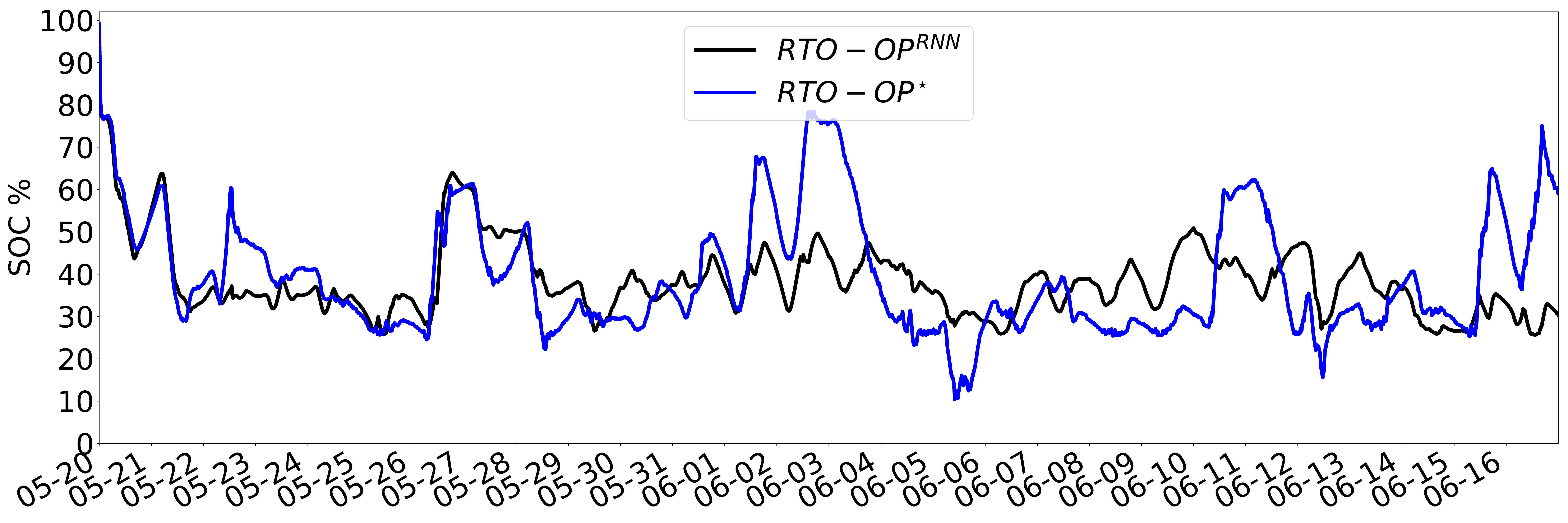} \\[1mm] 
	\includegraphics[width=0.9\linewidth]{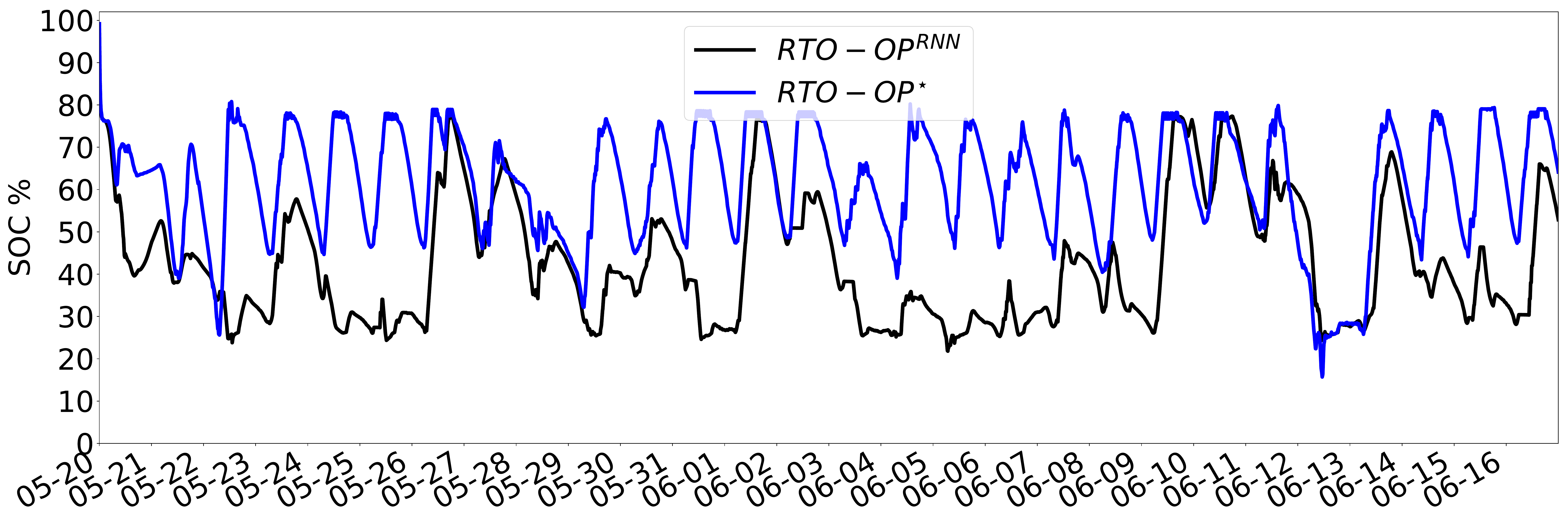}  
	\captionsetup{justification=centering}
	\caption{Case 1 (top) and 2 (bottom) SOC.}
	\label{fig:case21_soc_with_reserve}
\end{figure}

Table \ref{tab:results_w_reserve} provides the simulation results by taking into account the symmetric reserve. Figure \ref{fig:case3_soc_r0_vs_20} depicts on case 3 the behavior differences between $\mbox{RTO-OP}^{RNN}$ without and with symmetric reserve. 	Figures \ref{fig:case21_with_reserve_peak_costs} and \ref{fig:case21_soc_with_reserve} show the SOC and peaks costs evolution of case 2 \& 1.
The controller tends to maintain a storage level that allows $\mbox{RTO-OP}^{RNN}$ to better cope with forecast error. Indeed for case 3, there is no more peak reached by $\mbox{RTO-OP}^{RNN}$, only 1 kW for case 2 and it has been almost divided by two for case 1. However, this behavior tends to increase the energy cost if the PV production is important in comparison with the consumption, such as in case 3. Indeed, the controller will tend to store more energy in the battery instead of exporting it. 
$\mbox{RTO-OP}^{\star}$ did not perform better with the symmetric reserve. The symmetric reserve competes with the peak management and the $\mbox{RTO-OP}^{\star}$ tends to not discharge completely the battery even if it is required to avoid a peak. In case 2, the peak is reached on $\twelveofjune$ around 08:00. The controller could have avoided it by totally discharging the battery but did not maintain the reserve level. This is the same behavior in case 1 where the peak could have been limited if all the battery was discharged. There is an economic trade-off to reach to manage the peak and the reserve simultaneously depending on the valorization or not on the market of the symmetric reserve. The reserve can also be valorized internally to cope with non or difficult forecastable events such as a sudden drop of the export or import limits due to loss of equipment or grid congestion.
\begin{table}[tb]
	\begin{center}
		\scriptsize
		\renewcommand\arraystretch{1.5}
		\caption{Results with symmetric reserve.}
		\begin{tabular}{l|r|r|r|r|r|r}
			\hline  \hline 
			Case 1 & $c_E$ & $c_p$ & $c_t$ & $\Delta_p$ & $\mbox{I}_{tot}$ & $\mbox{E}_{tot}$ \\ \hline  
			$\mbox{RTO-OP}^{RNN}$	 & 10.50 & 2.12  & 12.62 & 53  & 65 &  3    \\ \hline 
			$\mbox{RTO-OP}^{\star}$	 & 10.47 & 2.75  & 13.22 & 69  & 65 & 2 \\ \hline  \hline 
			Case 2 & $c_E$ & $c_p$ & $c_t$ & $\Delta_p$ & $\mbox{I}_{tot}$ & $\mbox{E}_{tot}$ \\ \hline  
			$\mbox{RTO-OP}^{RNN}$	 & 5.33  & 0.04 & 5.37 & 1  & 41 & 27   \\ \hline 
			$\mbox{RTO-OP}^{\star}$	 & 4.78  & 0.99 &  5.77 & 25     & 35 & 20\\ \hline \hline 
			Case 3 & $c_E$ & $c_p$ & $c_t$ & $\Delta_p$ & $\mbox{I}_{tot}$ & $\mbox{E}_{tot}$ \\ \hline  
			$\mbox{RTO-OP}^{RNN}$	 & -0.04 & 0 & -0.04 & 0   & 24   & 99   \\ \hline 
			$\mbox{RTO-OP}^{\star}$	 & -0.15 & 0 & -0.15 & 0     & 23.2 & 98 
		\end{tabular}
		\label{tab:results_w_reserve}
	\end{center}
\end{table}


\section{Conclusion}\label{sec:conclusion}

A two-level value function-based approach was introduced as a solution method for a multi-resolution  microgrid optimization problem. The value function computed by the operational planner based on PV and consumption forecasts allows coping with the forecasting uncertainties. The real-time controller solves an  entire optimization problem including the future information propagated by the value function.
This approach has been tested on the MiRIS microgrid case study with PV and consumption data monitored on-site. The results demonstrate the efficiency of this method to manage the peak in comparison with a Rule-Based Controller. This test case is completely reproducible as all the data used are open, PV, consumption monitored and forecasted including the weather forecasts.
The proposed approach can be extended in several ways. The deterministic formulation of the operational planning problem could be extended to a stochastic formulation, to cope with probabilistic forecasts. Balancing market mechanisms could be introduced. Finally, the approach could be extended to a community by considering several entities inside the microgrid.

\section{Acknowledgment}
The authors would like to thank John Cockerill and Nethys for their financial support, and Xavier Fettweis of the Laboratory of Climatology of ULi\`ege who produced the weather forecasts based on the MAR regional climate model.

\bibliographystyle{IEEEtran}
\bibliography{biblio}

\newpage

\section{Notation}\label{sec:notation}

\subsection*{\textbf{Set and indices}}

\begin{itemize}
	\item $d$ index of a device
	\item $t$, $t'$ indexes of a RTO and OP time periods
	\item $\tau(t)$ beginning of the next market period at time $t$
	\item $\mathcal{T}_i(t) = \{t, t+\Delta t, ..., t+T_i\}$ set of RTO time periods
	\item $\mathcal{T}^m_a(t) = \{\tau(t), \tau(t)+\Delta \tau,..., \tau(t+T_a)\}$ set of OP time periods
	\item $T_a$, $T_l$ time durations, with $T_a \ll T_l$
	\item $\mathcal{D}^k$ set of non-flexible loads ($k = \text{nfl}$), sheddable loads ($k = \text{she}$), steerable generators ($k = \text{ste}$),  non-steerable generators ($k = \text{nst}$), storage devices  ($k = \text{sto}$)
\end{itemize}

\subsection*{\textbf{Parameters}}

\begin{itemize}
	\item $\Delta_t$ time delta between $t$ and the market period (minutes)
	\item $\Delta_\tau$ market period (minutes)
	\item $H_T$ forecasting horizon (hours)
	\item $\hat{\omega}$ forecast of a random vector $\omega$
	\item $\chargeEfficiency$, $\dischargeEfficiency$ charge and discharge efficiencies (\%)
	\item $\chargerate$, $\dischargerate$ maximum charging and discharging powers (kW)
	\item $\nonFlexible_{d,t}$ non-flexible power consumption (kW)
	\item $\sheddable_{d,t}$ flexible power consumption (kW)
	\item $\initialCharge_{d,t}$ initial state of charge of battery~$d$ (kWh)
	\item $p_h$ maximum peak over the last twelve months (kW)
	\item $\price{p}$ yearly peak power cost (\euro /kW)
	\item $\price{s}_{OP}$ unitary revenue for providing reserve (\euro /kW)
	\item $\price{s}_{RTO}$ unitary RTO symmetric reserve penalty (\euro /kW)
	\item $\pi^k_{d,t}$ cost of load shedding ($k = \text{she}$), generating energy ($k = \text{ste}$),  curtailing generation ($k = \text{nst}$) (\euro /kWh)
	\item $\BSSsFee_{d,t}$ fee to use the battery~$d$ (\euro /kWh)
	\item $\gridSalePrice_t$, $\gridBuyPrice_t$ energy prices of export and import  (\euro /kWh)
	\item $\gridBuyPrice_d$, $\gridBuyPrice_n$ energy prices of day and night  imports (\euro /kWh)
	\item $\maxImportFromGrid$, $\maxExportToGrid$ maximum import and export limits (kW) 
	\item $\mbox{PV}_p$, $\mbox{C}_p$ PV and consumption capacities (kW)
	\item $\mbox{S}_p$ storage capacity (kWh)
	\item $\maxcharge$, $\mincharge$ maximum and minimum battery capacities (kWh)
\end{itemize}

\subsection*{\textbf{Forecasted or computed variables}}

\begin{itemize}
	\item $a_t$ action at $t$
	\item $a^m_t$ purely market related actions
	\item $a^d_t$ set-points to the devices of the microgrid
	\item $a^k_{d,t} $ fraction of load shed ($k = \text{she}$), generation activated ($k = \text{ste}$),  generation curtailed ($k = \text{nst}$) ($ [0,1]$)
	\item $\charge_{d,t} $, $\discharge_{d,t} $ fraction of the maximum charging and discharging powers used for battery~$d$ ($ [0,1]$)
	\item $\exportGrid_t$, $\importGrid_t$ energy export and import (kWh)
	\item $\delta p_{t'}$ OP peak difference between peak at~$t'$ and $p_h$ (kW) 
	\item $\delta p_{\tau(t-\Delta_\tau),\tau(t)}$ RTO peak difference between peak at~$\tau(t)$ and $p_h$ (kW) 
	\item $s^{TSO}_{t}$ TSO symmetric reserve signal ($0;1$)
	\item $\reserve{sym}$ symmetric reserve (kW)
	\item $\Delta \reserve{sym}$ reserve difference between OP and RTO (kW) 
	\item $\reserveBSSInc$, $\reserveBSSDec$ upward and downward reserves of power available and provided by storage device~$d$ (kW)
	\item $\SOC_{d,t}$ state of charge of battery~$d$ (kWh)
	\item $s_t$ microgrid state at time $t$
	\item $s^m_t$ information related to the current market position
	\item $s^d_t$ state of the devices
	\item $v_t$ the cost-to-go function
	\item $\hat{\omega}$ forecast of a random vector $\omega$
	\item $\overline{\mbox{X}}$ average of a variable $X$ (kW)
	\item $\mbox{X}_{max}$, $\mbox{X}_{min}$ maximum and minimum of $X$ (kW)
	\item $\mbox{X}_{std}$ standard deviation of $X$ (kW)
	\item $c_E$, $c_p$, $c_t$ energy, peak and total costs (k\euro)
	\item $\Delta_p$ peak increment (kW)
	\item $\mbox{I}_{tot}$, $\mbox{E}_{tot}$ total import and export (MWh)
\end{itemize}

\section{Annex: forecasting methodology}
\label{sec:forecasting}

The inputs of the forecasting method are historical and external data, a forecasting horizon $H_T$, a resolution, and a forecast frequency. The outputs are the PV production and the consumption. In this study, the input data are weather forecasts and past PV production and consumption series. The horizon is the time range of the forecasts from a few hours to several hours or days. The resolution is the time discretization of the forecast from a few minutes to several hours. The forecast frequency indicates the periodicity at which the forecasts are computed. For instance, a forecasting module with $H_T=24$ hours, a resolution and periodicity of 15 minutes, computes each quarter, a quarterly forecast for the twenty-four hours ahead. This paper focuses on the real-time control of microgrids based on planning that requires a forecast horizon of a few hours up to a few days.

Two "classic" deterministic techniques are implemented, a Recurrent Neural Network (RNN) with the Keras Python library \cite{chollet2015keras} and a Gradient Boosting Regression (GBR) with the Scikit-learn Python library \cite{scikit-learn}. The RNN is a Long Short Term Memory (LSTM) with one hidden layer composed of $2 \times n+1$ neurons with $n$ the number of input features. Both techniques are implemented with a Multi-Input Multi-Output (MIMO) approach \cite{taieb2012review}. The MIMO strategy consists of learning only one model, $\widehat{m}$, as follows
\begin{equation}
[\widehat{w}^{\tau_1}, ..., \widehat{w}^{\tau_{H_T}}] = \widehat{m} \bigg[ w^{\tau_0}, ..., w^{\tau_{-4}}, \widehat{w}^{\tau_1}_i, ..., \widehat{w}^{\tau_{H_T}}_i \bigg ].
\end{equation}
With $\widehat{w}$ the variable to forecast (PV, consumption, etc), $\widehat{w}_i$ the forecast of the $i^{\text{th}}$ weather variable such as direct solar irradiance, wind speed, air ambient temperature, etc. 
The forecast is computed each quarter and composed of $H_T / \Delta \tau$ values $[\widehat{y}^{\tau_1}, ..., \widehat{y}^{\tau_{H_T}}]$.  In our case study, $H_T / \Delta_\tau = 96$ with $H_T$ = 24 h and $\Delta_\tau$ = 15 minutes. The forecasting process is implemented as a rolling forecast methodology. The Learning Set (LS) is refreshed every six hours. The LS is limited to the week preceding the forecasts, to maintain a reasonable computation time.

\begin{figure}[tb]
	\centering
	\includegraphics[width=1\linewidth]{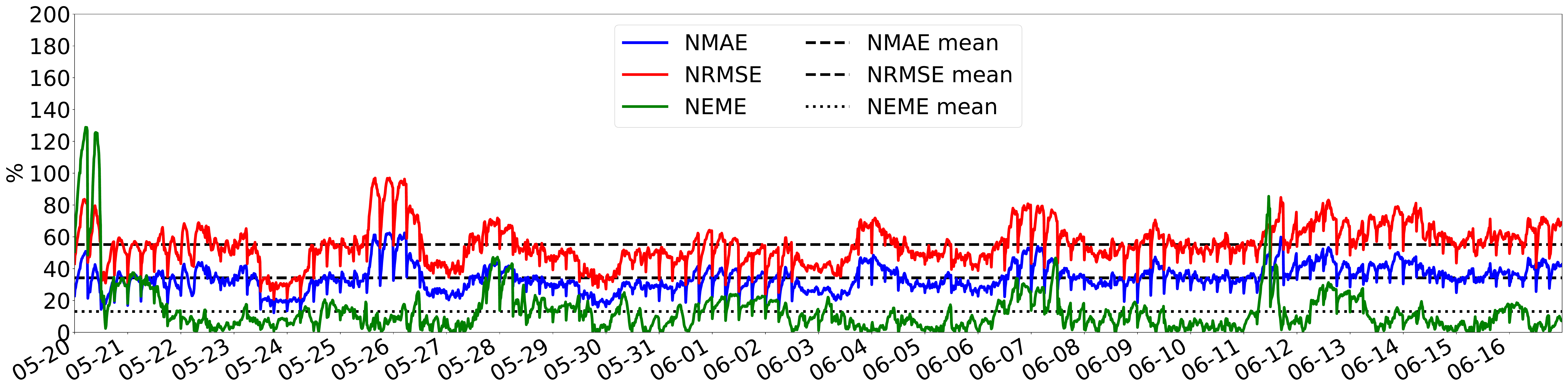} 
	\includegraphics[width=1\linewidth]{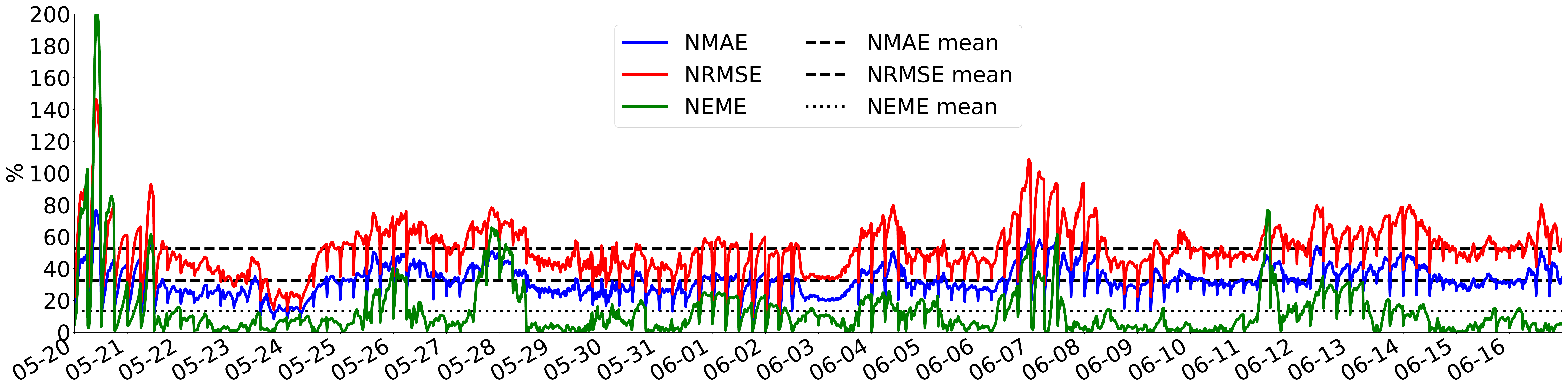} 
	\captionsetup{justification=centering}
	\caption{PV forecast scores for GBR (top) and RNN (bottom).}
	\label{fig:pvforecast_scores}
	\centering
	\includegraphics[width=1\linewidth]{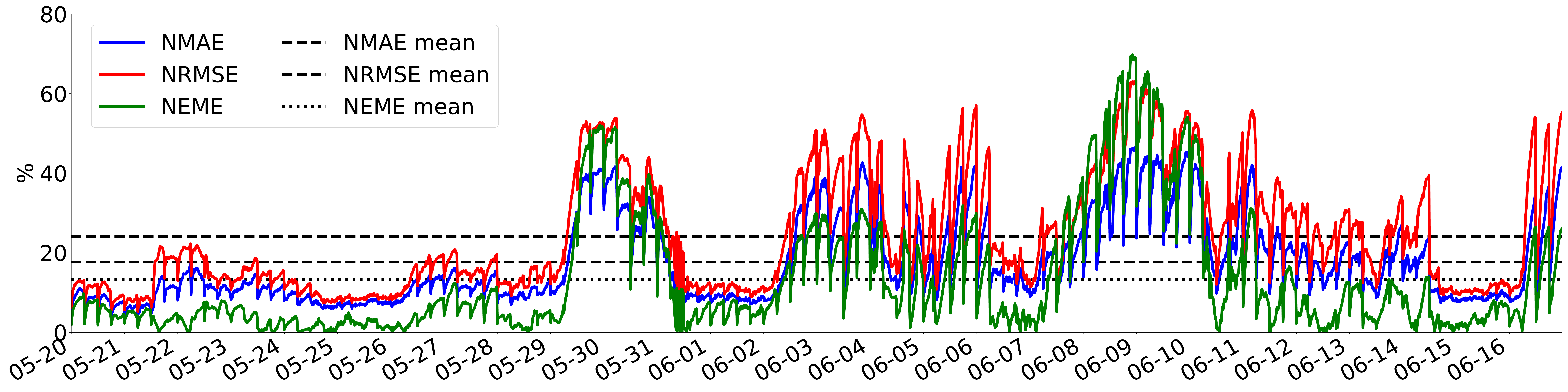} 
	\includegraphics[width=1\linewidth]{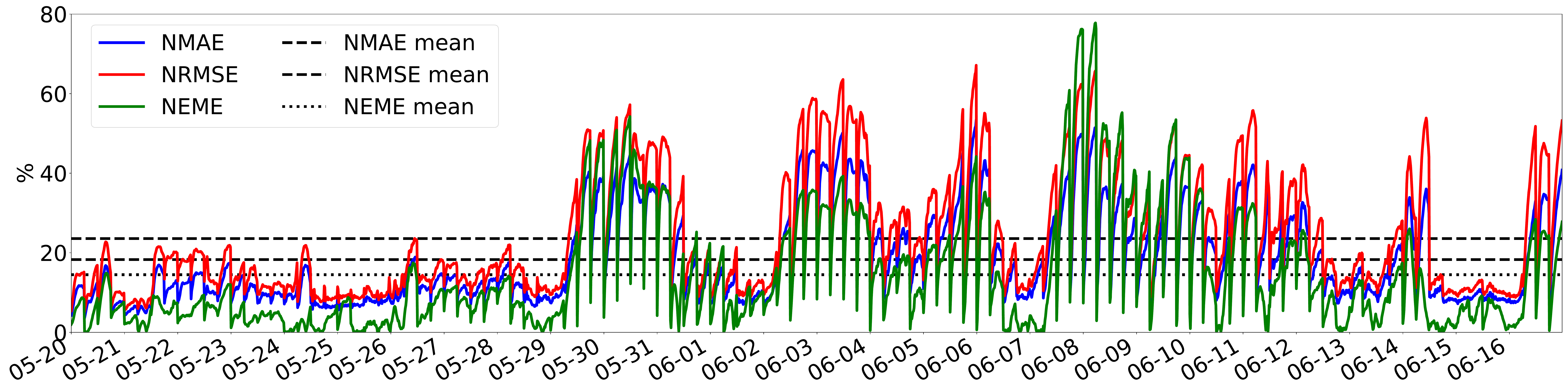} 
	\captionsetup{justification=centering}
	\caption{Consumption forecast scores for GBR (top) and RNN (bottom).}
	\label{fig:consoforecast_scores}
\end{figure}

The forecasts are evaluated using three deterministic metrics: the Normalized Mean Absolute Error (NMAE), the Normalized Root Mean Squared Error (NRMSE), and the Normalized Energy Measurement Error (NEME). The NEME is an NMAE of the energy summed over the entire forecasting horizon. The mean scores $NMAE_{H_T}$,  $NRMSE_{H_T}$ and $NEME_{H_T}$ for a forecasting horizon $H_T$ are computed over the entire simulation data set. The normalizing coefficient for computing the NMAE and the NRMSE is the mean of the absolute value of the PV and consumption over all the simulation data set. Figures \ref{fig:pvforecast_scores} and \ref{fig:consoforecast_scores} provide the scores for both GBR and RNN techniques computed for each quarter of the simulation data set.

\end{document}